\documentclass{emulateapj}
\usepackage{color}

\def\simlt{\lower.5ex\hbox{$\; \buildrel < \over \sim \;$}}
\def\simgt{\lower.5ex\hbox{$\; \buildrel > \over \sim \;$}}

\def\gsim{\lower 2pt \hbox{$\, \buildrel {\scriptstyle >}\over
{\scriptstyle \sim}\,$}}
\def\lsim{\lower 2pt \hbox{$\, \buildrel {\scriptstyle <}\over
{\scriptstyle \sim}\,$}}

\def\deg{\ifmmode ^{\circ}
         \else $^{\circ}$\fi}
\def\pdeg{\ifmmode
           $\setbox0=\hbox{$^{\circ}$}\rlap{\hskip.11\wd0 .}$^{\circ}
     \else \setbox0=\hbox{$^{\circ}$}\rlap{\hskip.11\wd0 .}$^{\circ}$\fi}
     
\def\adf{\ifmmode a_\mathrm{df} \else $a_\mathrm{df}$ \fi}
\def\rperi{\ifmmode r_\mathrm{p} \else $r_\mathrm{p}$ \fi}
     
\def\vkick{\ifmmode v_\mathrm{k} \else $v_\mathrm{k}$ \fi}
\def\rkick{\ifmmode r_\mathrm{k} \else $r_\mathrm{k}$ \fi}
\def\rinfl{\ifmmode r_\mathrm{infl} \else $r_\mathrm{infl}$ \fi}
\def\vesc{\ifmmode v_\mathrm{esc} \else $v_\mathrm{esc}$ \fi}
\def\vrot{\ifmmode v_\mathrm{rot} \else $v_\mathrm{rot}$ \fi}
\def\vrel{\ifmmode v_\mathrm{rel} \else $v_\mathrm{rel}$ \fi}
\def\tcoll{\ifmmode t_\mathrm{coll} \else $t_\mathrm{coll}$ \fi}
     
\def\pc{\ifmmode \mathrm{pc} \else $\mathrm{pc}$ \fi}
\def\mpc{\ifmmode \mathrm{Mpc} \else $\mathrm{Mpc}$\fi}
\def\mpcthree{\ifmmode \mathrm{Mpc}^{-3} \else $\mathrm{Mpc}^{-3}$\fi}
\def\gpcthree{\ifmmode \mathrm{Gpc}^{-3} \else $\mathrm{Gpc}^{-3}$\fi}

\def\kelvin{\ifmmode \mathrm{K} \else {$\mathrm{K}$}\fi}
\def\kev{\ifmmode \mathrm{keV} \else $\mathrm{keV}$ \fi}

\def\lsun{\ifmmode {L_\odot} \else $L_\odot$\fi}
\def\msun{\ifmmode M_\odot \else $M_\odot$\fi}
\def\msunyr{\ifmmode M_\odot~\mathrm{yr}^{-1} \else $M_\odot~\mathrm{yr}^{-1}$\fi}

\def\cosi{\ifmmode {\cos\,i} \else $\cos\,i$\fi}

\def\heii{\ifmmode {\rm He{\sc ii}} \else He~{\sc ii}\fi}
\def\mgii{\ifmmode {\rm Mg{\sc ii}} \else Mg~{\sc ii}\fi}
\def\ciii{\ifmmode {\rm C{\sc iii}]} \else C~{\sc iii}]\fi}
\def\civ{\ifmmode {\rm C{\sc iv}} \else C~{\sc iv}\fi}
\def\mgii{\ifmmode {\rm Mg{\sc ii}} \else Mg~{\sc ii}\fi}

\def\teff{\ifmmode {T_{\rm eff}} \else $T_{\rm eff}$\fi}
\def\tmax{\ifmmode {T_{\rm max}} \else $T_{\rm max}$\fi}

\def\mbh{\ifmmode {M_{\rm BH}} \else $M_{\rm BH}$\fi}
\def\led{\ifmmode L_{\mathrm{Ed}} \else $L_{\mathrm{Ed}}$\fi}
\def\lbolflare{\ifmmode L_{\mathrm{bol,flare}} \else $L_{\mathrm{bol,flare}}$\fi}
\def\lagn{\ifmmode L_{\mathrm{agn}} \else $L_{\mathrm{agn}}$\fi}
\def\lbolagn{\ifmmode L_{\mathrm{bol,agn}} \else $L_{\mathrm{bol,agn}}$\fi}
\def\lbol{\ifmmode L_{\mathrm{bol}} \else $L_{\mathrm{bol}}$\fi}
\def\mdot{\ifmmode {\dot M} \else $\dot M$\fi}
\def\mdoto{\ifmmode {\dot{M}_0} \else  $\dot{M}_0$\fi}
\def\mdotf{\ifmmode {\dot{M}_\mathrm{flare}} \else  $\dot{M}_\mathrm{flare}$\fi}

\def\hnot{\ifmmode H_0 \else H$_0$ \fi}

\def\vkep{\ifmmode v_\mathrm{Kep} \else $v_\mathrm{Kep}$ \fi}
\def\vc{\ifmmode v_\mathrm{c} \else $v_\mathrm{c}$ \fi}
\def\vctwo{\ifmmode v_\mathrm{c,250} \else $v_\mathrm{c,250}$ \fi}

\def\vthree{\ifmmode v_{1000} \else $v_{1000}$ \fi}
\def\vrel{\ifmmode v_\mathrm{rel} \else $v_\mathrm{rel}$ \fi}
\def\vkick{\ifmmode v_\mathrm{kick} \else $v_\mathrm{kick}$ \fi}
\def\vkickz{\ifmmode v_{\mathrm{kick},z} \else $v_{\mathrm{kick},z} $ \fi}
\def\vkicky{\ifmmode v_{\mathrm{kick},y} \else $v_{\mathrm{kick},y} $ \fi}
\def\vchar{\ifmmode v_\mathrm{char} \else $v_\mathrm{char}$ \fi}
\def\eflare{\ifmmode E_\mathrm{flare} \else $E_\mathrm{flare}$ \fi}
\def\ekick{\ifmmode E_\mathrm{kick} \else $E_\mathrm{kick}$ \fi}
\def\ecoll{\ifmmode E_\mathrm{coll} \else $E_\mathrm{coll}$ \fi}
\def\ezero{\ifmmode E_\mathrm{0} \else $E_\mathrm{0}$ \fi}
\def\efac{\ifmmode \xi_\mathrm{E} \else $\xi_\mathrm{E}$ \fi}
\def\tqso{\ifmmode t_\mathrm{QSO} \else $t_\mathrm{QSO}$ \fi}
\def\tflare{\ifmmode t_\mathrm{flare} \else $t_\mathrm{flare}$ \fi}
\def\tzero{\ifmmode t_\mathrm{0} \else $t_\mathrm{0}$ \fi}
\def\tfac{\ifmmode \xi_\mathrm{t} \else $\xi_\mathrm{t}$ \fi}
\def\gfac{\ifmmode f_\mathrm{g} \else $f_\mathrm{g}$ \fi}
\def\lflare{\ifmmode L_\mathrm{flare} \else $L_\mathrm{flare}$ \fi}
\def\fflare{\ifmmode F_\mathrm{flare} \else $F_\mathrm{flare}$ \fi}
\def\nflare{\ifmmode N_\mathrm{flare} \else $N_\mathrm{flare}$ \fi}
\def\tshock{\ifmmode T_\mathrm{shock} \else $T_\mathrm{shock}$ \fi}
\def\rmin{\ifmmode R_\mathrm{1} \else $R_\mathrm{1}$ \fi}
\def\rmax{\ifmmode R_\mathrm{2} \else $R_\mathrm{2}$ \fi}
\def\rbound{\ifmmode R_\mathrm{b} \else $R_\mathrm{b}$ \fi}
\def\pbound{\ifmmode P_\mathrm{b} \else $P_\mathrm{b}$ \fi}
\def\mbound{\ifmmode M_\mathrm{b} \else $M_\mathrm{b}$ \fi}
\def\mbo{\ifmmode M_{\mathrm{b}0} \else $M_{\mathrm{b}0} $ \fi}
\def\ebo{\ifmmode E_{\mathrm{b}0} \else $E_{\mathrm{b}0} $ \fi}
\def\efinal{\ifmmode E_\mathrm{final} \else $E_\mathrm{final} $ \fi}
\def\tbound{\ifmmode t_\mathrm{b} \else $t_\mathrm{b}$ \fi}
\def\tagn{\ifmmode t_\mathrm{AGN} \else $t_\mathrm{AGN}$ \fi}
\def\rlim{\ifmmode R_\mathrm{lim} \else $R_\mathrm{lim}$ \fi}
\def\vlim{\ifmmode v_\mathrm{lim} \else $v_\mathrm{lim}$ \fi}
\def\vphi{\ifmmode v_\phi \else $v_\phi$ \fi}
\def\mlim{\ifmmode M_\mathrm{lim} \else $M_\mathrm{lim}$ \fi}
\def\tlim{\ifmmode t_\mathrm{lim} \else $t_\mathrm{lim}$ \fi}
\def\llim{\ifmmode L_\mathrm{lim} \else $L_\mathrm{lim}$ \fi}
\def\fqso{\ifmmode f_\mathrm{QSO} \else $f_\mathrm{QSO}$ \fi}

\def\hbeta{\ifmmode \rm{H}\beta \else H$\beta$\fi}
\def\hbetan{\ifmmode \rm{H}\beta_{\rm n} \else H$\beta_{\rm n}$\fi}

\def\halpha{\ifmmode \rm{H}\alpha \else H$\alpha$\fi}
\def\lalpha{\ifmmode \rm{Ly}\alpha \else Ly$\alpha$}

\def\dvhb{\ifmmode \Delta v_{\hbeta} \else $\Delta v_{\hbeta}$\fi}
\def\dvmg{\ifmmode \Delta v_{\rm{Mg}} \else $\Delta v_{\rm{Mg}}$\fi}

\def\muobs{\ifmmode {\mu_{o}} \else  $\mu_{o}$ \fi}
\def\cosi{\ifmmode {\mathrm{cos}\,i} \else $\mathrm{cos}\,i$\fi}

\def\teff{\ifmmode {T_{eff}} \else $T_{eff}$ \fi}
\def\tmax{\ifmmode {T_{max}} \else $T_{max}$ \fi}

\def\tauh{\ifmmode {\tau_{\rm H}} \else $\tau_{\rm H}$ \fi}

\def\yr{\ifmmode {\rm yr} \else  yr \fi}
\def\kpc{\ifmmode {\rm kpc} \else  kpc \fi}
\def\Mpc{\ifmmode {\rm Mpc} \else  Mpc \fi}
\def\yr{\ifmmode {\rm yr} \else  yr \fi}
\def\kms{\ifmmode \rm km~s^{-1}\else $\rm km~s^{-1}$\fi}
\def\cm{\ifmmode {\rm cm} \else  cm \fi}
\def\cmmitwo{\ifmmode \rm cm^{-2} \else $\rm cm^{-2}$\fi}
\def\cmmithree{\ifmmode \rm cm^{-3} \else $\rm cm^{-3}$\fi}
\def\cmps{\ifmmode \rm cm~s^{-1}\else $\rm cm~s^{-1}$\fi}
\def\cmpsps{\ifmmode \rm cm~s^{-2}\else $\rm cm~s^{-2}$\fi}
\def\kmps{\ifmmode \rm km~s^{-1}\else $\rm km~s^{-1}$\fi}
\def\kmpspmpc{\ifmmode \rm km~s^{-1}~Mpc^{-1} \else
    $\rm km~s^{-1}~Mpc^{-1}$\fi}
  
\def\gcmthree{\ifmmode \rm g~cm^{-3} \else $\rm g~cm^{-3}$\fi}
\def\gcmtwo{\ifmmode \rm g~cm^{-2} \else $\rm g~cm^{-2}$\fi}
   
\def\erg{\ifmmode {\rm erg} \else $\rm erg$ \fi}
\def\ergps{\ifmmode {\rm erg~s^{-1}} \else $\rm erg~s^{-1}$ \fi}
\def\ergcms{\ifmmode \rm erg~cm^{-2}~s^{-1} \else $\rm erg~cm^{-2}~s^{-1}$ \fi}
\def\ergcmshz{\ifmmode \rm erg~s^{-1}~cm^{-2}~Hz^{-1} \else $\rm
erg~cm^{-2}~s^{-1}~Hz^{-1}$ \fi}
\def\ergcmsa{\ifmmode \rm erg~cm^{-2}~s^{-1}~\AA^{-1} \else $\rm
erg~cm^{-2}~s^{-1}~\AA^{-1}$ \fi}
\def\ergshz{\ifmmode \rm erg s^{-1} Hz^{-1} \else
   $\rm erg s^{-1} Hz^{-1}$ \fi}

\def\lam{\ifmmode {\lambda} \else {$\lambda$} \fi}
\def\llam{\ifmmode {L_\lambda} \else  $L_\lambda$ \fi}
\def\lamLlam{\ifmmode \lambda L_{\lambda}(5100) \else {$\lambda L_{\lambda}(5100)$} \fi}
\def\nuLnu{\ifmmode \nu L_{\nu}(5100) \else {$\nu L_{\nu}(5100)$} \fi}
\def\ilam{\ifmmode {I_\lambda} \else  $I_\lambda$ \fi}
\def\flam{\ifmmode {F_\lambda} \else  $F_\lambda$ \fi}
\def\inu{\ifmmode {I_\nu} \else  $I_\nu$ \fi}
\def\fnu{\ifmmode {F_\nu} \else  $F_\nu$ \fi}
\def\bnu{\ifmmode {B_\nu} \else  $B_\nu$ \fi}

\def\msigma{\ifmmode M_{\sigma} \else $M_{\sigma}$\fi}
\def\mbulge{\ifmmode M_{\mathrm{bulge}} \else $M_{\mathrm{bulge}}$\fi}
\def\mgal{\ifmmode M_{\mathrm{gal}} \else $M_{\mathrm{gal}}$\fi}
\def\lgal{\ifmmode L_{\mathrm{gal}} \else $L_{\mathrm{gal}}$\fi}
\def\lbulge{\ifmmode L_{\mathrm{bulge}} \else $L_{\mathrm{bulge}}$\fi}
\def\mgalstar{\ifmmode M^*_{\mathrm{gal}} \else $M^*_{\mathrm{gal}}$\fi}
\def\mbhsig{$M_{\mathrm{BH}} - \sigma$}
\def\mbhsigstar{\ifmmode M_{\mathrm{BH}} - \sigma_* \else $M_{\mathrm{BH}} - \sigma_*$ \fi}
\def\mbhlum{\ifmmode M_{\mathrm{BH}} - L \else $M_{\mathrm{BH}} - L$ \fi}
\def\deltalogmbh{\ifmmode \Delta~{\mathrm{log}}~M_{\mathrm{BH}} \else $\Delta$~log~$M_{\mathrm{BH}}$\fi}

\def\sigstar{\ifmmode \sigma_* \else $\sigma_*$\fi}
\def\sigthree{\ifmmode \sigma_{\mathrm{[O~III]}} \else $\sigma_{\mathrm{[O~III]}}$\fi}
\def\sigtwo{\ifmmode \sigma_{\mathrm{[O~II]}} \else $\sigma_{\mathrm{[O~II]}}$\fi}
\def\signl{\ifmmode \sigma_{\mathrm{NL}} \else $\sigma_{\mathrm{NL}}$\fi}
\def\wthree{\ifmmode {\rm FWHM({[O~III]})} \else $FWHM({[O~III]})$ \fi}
\def\wtwo{\ifmmode {\rm FWHM({[O~II]})} \else $FWHM({[O~II]})$ \fi}
\def\mthree{\ifmmode M_{\mathrm [O~III]} \else $M_{\mathrm [O~III]}$ \fi}
\def\mtwo{\ifmmode M_{\mathrm [O II]} \else $M_{\mathrm [O II]}$ \fi}

\def\lbreak{\ifmmode L_{\mathrm{break}} \else $L_{\mathrm{break}}$\fi}
\def\lcut{\ifmmode L_{\mathrm{cut}} \else $L_{\mathrm{cut}}$\fi}

\received{}
\accepted{}

\slugcomment{Submitted to ApJ}

\shortauthors{Shields \& Bonning}
\shorttitle{Runaway Black Hole}

\begin{document}

\title{A Captured Runaway Black Hole in NGC 1277?}

\author{G.~A. Shields\altaffilmark{1}, E.~W. Bonning \altaffilmark{2}}

\altaffiltext{1}{Department of Astronomy, University of Texas, Austin,
TX 78712; shields@astro.as.utexas.edu} 

\altaffiltext{2}{Quest University Canada, 3200 University Boulevard,
Squamish, BC, V8B 0N8, Canada; erin.bonning@questu.ca} 

\begin{abstract}

Recent results indicate that the compact lenticular galaxy NGC 1277 in the Perseus Cluster 
contains a black hole of mass $\sim10^{10}~\msun$.
This far exceeds the expected mass of the central black hole in a galaxy of the
modest dimensions of NGC 1277.  We suggest that this giant black hole was ejected from the nearby
giant galaxy NGC 1275 and subsequently captured by NGC 1277.  The ejection was the
result of gravitational radiation recoil when two large black holes merged following
the merger of two giant ellipticals that helped to form NGC 1275.    The  black hole wandered
in the cluster core until it was captured in a close encounter with NGC 1277.  The
migration of black holes in clusters may be a common occurrence.

\end{abstract}


\section{Introduction}
\label{sec:intro}

Recent imaging and spectroscopic analysis of the compact lenticular galaxy NGC 1277,
located in the Perseus Cluster,
indicates a central ultra-massive black hole (UMBH) 
with a mass of $\mbh \approx 1.7\times10^{10}~\msun$ \citep[``VB12'']{vdbosch12}.
  The mass exceeds
by two orders of magnitude the value expected on the basis of the galaxy's  luminosity.
 In fact, it is one of the largest black hole masses reported to date on the basis of stellar dynamics.  
The origin of this black hole is therefore of great interest.
Such a massive black hole might
be expected to form in the center of a giant elliptical galaxy of the kind found in the
centers of rich clusters of galaxies (McConnell et al. 2012).   
Here we propose that the giant black hole in NGC~1277 did indeed originate in another,
much larger galaxy in the cluster.   Its formative event was the merger of two
giant elliptical galaxies, each having a massive black hole similar to
 the $\sim10^{9.8}~\msun$ black hole in M87 \citep{gebhardt11}.  
 The in-spiral and merger of these holes
 resulted in ejection of the product black hole from the merged host galaxy by means
 of gravitational radiation recoil. 
 We identify the progenitor galaxy with the giant cD galaxy NGC 1275 
that dominates the cluster.   The ejected black hole wandered in the core of the cluster until
 a chance encounter with NGC 1277 led to
its capture and orbital decay into the nucleus.
Meanwhile, NGC 1275 reformed its present, relatively small black hole 
through subsequent mergers and accretion of gas.

\section{The Escape}
\label{sec:escape}
 
On the basis of
dynamical models of the stellar light profile and kinematics of NGC 1277,
VB12 derived a black hole mass of $(17\pm3)\times10^9~\msun$.
This large mass is out of proportion to the host galaxy, for which
VB12 give a total stellar mass of $(1.2\pm0.4)\times10^{11}~\msun$.   These authors
fit the light profile with a disky component plus a central pseudo-bulge containing
24\%\ of the light.   Therefore
\mbh\ is 14\%\ of the total stellar mass and $~\sim59\%$ of the bulge mass.
In contrast, black holes in galactic nuclei typically have a mass $\sim10^{-2.9}$ of the bulge mass
\citep{kormendy01}.    Although NGC 1277 has an unusually large stellar velocity dispersion
for its luminosity, $\sigstar \approx 333~\kms$, its black hole mass still exceeds by
nearly an order-of-magnitude the value expected by the normal \mbhsig\ relationship (VB12).
 NGC 1277 is such an extreme outlier in the \mbhsig\ plane as to raise the question of
a qualitatively different evolutionary history.

Black holes with enormous masses similar to that in NGC 1277 have been discovered in recent years.
They are mostly found in large elliptical galaxies, often brightest cluster galaxies (BCGs).
\citet{mcconnell12} present measurements for four BCGs and summarize
earlier work.   Notable cases include 
$\mbh = (21\pm16)\times10^9~\msun$ for NGC 4889 in the Coma cluster,
$(9.7\pm2.5)\times10^9~\msun$ for NGC 3842 in Abell 1367,  
$(3.6\pm1.1)\times10^9~\msun$ for NGC 6086 in Abell 2162, and
$(6.6\pm0.4)\times10^9~\msun$ for M87 in the Virgo Cluster 
\citep{mcconnell12, gebhardt11}.
Consistent with these measurements, 
cosmological simulations by \citet{yoo07} show that black holes with mass up to $\sim1.5\times10^{10}~\msun$
can form by mergers in massive clusters.

NGC 1277 is located in the core of
the Perseus Cluster of galaxies ($z = 0.018$), one of the largest nearby clusters (richness class 2).  
The dominant galaxy of this cluster is NGC 1275, a large cD galaxy with a radio source
(Per A), X-ray emission  \citep[and references therein]{fabian11}, and optical AGN activity with a narrow-line (Sy~2) spectrum 
\citep{seyfert43}. 
The nucleus of NGC 1275 is the most natural place to form a UMBH (followed by NGC 1272, the next brightest galaxy in the cluster).
For the largest galaxies, the bulge luminosity is a better predictor of \mbh\  than is \sigstar\ \citep[e.g.,][]{lauer07a}.
The luminosity of NGC 1275 is half a magnitude fainter (in $M_{\mathrm V}$) than NGC 4889, and similar to that of
NGC 3842, according to the Hyperleda database \citep{paturel03}.\footnote{http://leda.univ-lyon1.fr}
Thus, it is reasonable to consider the possibility that 
a UMBH of the mass of the one in NGC 1277 may have originally formed in NGC 1275.

How might a UMBH have been ejected from the nucleus of NGC 1275?  One possibility is
 gravitational radiation recoil when two black holes merge \citep{merritt04}.   The escape velocity
from the nucleus may be estimated as  $\vesc \approx 5\sigstar$ \citep{merritt09},
giving   $\vesc  \approx 1250~\kms$ based on $\sigstar \approx 250~\kms$ for NGC 1275 \citep{heckman85}.
Gravitational radiation recoil during the final merger of spinning black holes
is capable of launching the product black hole with kick velocities
upwards of several thousand $\kms$  \citep{campa07a,gonzalez07}. The magnitude and
probability of the recoil velocity is dependent on the spin alignment of
the black holes. Initially, kicks of up to 4000~$\kms$ were predicted
for anti-aligned spins in the orbital plane \citep{campa07b}. This may be an astrophysically disfavored
scenario in the case of gas-rich mergers, as accretion may align the black hole spins with the
binary orbital axis, limiting recoil
velocities to several hundred $\kms$\citep{bogda07, dotti10}.
 In recent years, further exploration of non-linear
spin couplings has indicated that even larger kicks can result from BH
spins partially aligned with the orbital angular momentum 
\citep{loustoZ11,lousto12a,lousto13}.
The probability of a large recoil velocity increases in the light of these results.  For both black holes maximally spinning,
\citet{lousto13} give a maximum kick of $4900~\kms$ for equal mass holes, dropping only to $4500~\kms$ for
$q \equiv M_2/M_1 = 0.5$.   The probability is  9\% for kicks greater than $1000~\kms$ 
in the ``hot accretion''  cosmology-based simulations by \citet{lousto13}.  However, this reflects the effect of
accretion of gas in aligning the black hole spins and suppressing large kicks.   The merger leading to the formation
of a UMBH in NGC 1275 may well have been ``dry'', in which case the simulations with random spin orientations
by \citet{loustoNZC11} may be more appropriate.    Figure 26 of \citet{loustoNZC11} indicates a fraction $\sim25\%$ of mergers with nearly
equal masses will give $\vkick > 1250~\kms$.  Even this value may be pessimistic, because it assumes a uniform distribution of
spin magnitudes, whereas astrophysical black holes are likely to be rapidly spinning because of past accretion and mergers.
Furthermore, this value does not reflect the increase in \vkick\ caused by the new ``cross kick'' and ``hangup kick'' effects
discussed by \citet{lousto13}, and references therein.  Since large kicks can occur for substantially unequal black hole masses, 
there may be more than one opportunity for a merger and black hole ejection as a BCG grows.  Accordingly, the production of runaway black holes may be a common occurrence in clusters and groups of galaxies.  

The escaping black hole will carry with it a compact cluster of bound stars \citep{merritt09}.   Essentially,
this will involve the stars that were within the  radius such that they are bound to the hole after the kick, $\rkick =  G \mbh/\vkick^2$.  If we follow Merritt et al. in assuming a power-law stellar density profile inside \rinfl\
such that $\rho \propto r^{-\gamma}$ then the mass of bound stars as a fraction of the black hole mass is
$f_\mathrm{b} \equiv M_\mathrm{b}/\mbh \approx 10^{-2}$ or less for $\gamma$ in the range 1 to 2.  This is consistent with
the more detailed treatment by Merritt et al.  Thus, the bound cluster, while an interesting potential diagnostic of the
kick velocity in other contexts, will be small in comparison to the observed stellar mass of NGC 1277.  The runaway
black hole must have merged with one or more galaxies to form the system that we observe as NGC 1277 today.

\section{The Capture}
\label{sec:capture}

For $\vesc = 1250~\kms$ from the nucleus of NGC 1275, the runaway UMBH will leave the galaxy with
a terminal velocity of $800~\kms$ for $\vkick = 1500~\kms$ or $1300~\kms$ for $\vkick = 1800~\kms$.  For comparison,
the velocity dispersion of the Perseus Cluster is $\sigma_\mathrm{cl} \approx 1300~\kms$ 
\citep{struble91}.
However, NGC 1277 is close to NGC 1275 both in position on the sky (80~kpc) and in line-of-sight velocity, 
with $\Delta v \equiv v_{1277} - v_{1275} = -280~\kms$ \citep{brunzendorf99}.    It appears to be part of an inner core or subcluster of galaxies encompassing NGC 1275 and NGC 1272.
We focus here on the hypothesis that the runaway black hole orbited in the vicinity of this subcluster until captured by
NGC 1277.  Based on inspection of images of the cluster, we approximate the subcluster with an area extending $\pm 0.1$~ arc min ($\pm 120~\kpc$) in R.A. and in declination from a center at   $\alpha = 49.9125,  \delta = +41.5278$.  
The catalog of \citet{brunzendorf99} gives 10 (14) galaxies in this region that are no fainter than 1~magn (2~magn) fainter than NGC 1277.
 The velocity dispersion of the 10 galaxies is $\sigma \approx 1100 ~\kms$.   
  Of these 10 galaxies, three including NGC 1277 have radial velocities differing by less than $300~\kms$ from NGC 1275.  In addition, PGC~12443,  only 1.1~magn fainter than NGC~1277, has $\Delta v = 222~\kms$.

A galaxy can capture the black hole in a close encounter if dynamical friction 
on the galaxy's stellar and dark matter background density $\rho$ robs the hole of enough orbital energy to leave it bound to the galaxy.
For supersonic velocities ($v >> \sigstar$), the Chandrasekhar dynamical friction formula can be expressed 
$\adf = -4\pi G^2 M \rho\, \mathrm{ln}\Lambda\,v^{-2}$,
where \adf\ is the deceleration, $v$ is the relative velocity and $\mathrm{ln}\Lambda \approx 6$ \citep{binney08}.  
VB12 find a relatively flat rotation curve with circular velocity $\vc \approx 250~\kms$.  For a rough estimate of the
dynamical friction efficiency, we therefore take
$\rho = v_\mathrm{c}^2/(4\pi G r^2)$ with $v_\mathrm{c} \approx \mathrm{const}$,
by analogy to a singular isothermal sphere (Binney \& Tremaine).
Then we have
$\adf  \approx (10^{-9.28}~\cmpsps)  M_{10} \vctwo^2 r_{10}^{-2} v_3^{-2}$,
where $M_{10} \equiv \mbh/(10^{10}~\msun)$, $ \vctwo \equiv \vc/(250~\kms)$ , $ r_{10} \equiv r/(10~\kpc)$, and $ v_3 = v/(10^3~\kms)$
is the encounter velocity.
We may roughly estimate the energy per unit mass lost to dynamical friction during the encounter as
$\Delta E \approx -\pi \adf(\rperi) \rperi$,
where \rperi\ is the distance of closest approach, and the factor $\pi$ is motivated by a straight line encounter at constant velocity.   
A more detailed calculation would take account of gravitational focusing giving $\rperi < b$, where $b$ is the impact parameter.  This can give \rperi\ several times smaller than $b$ for parameters of interest.   However,
the greater background density near \rperi\ is offset by the smaller radius and the higher velocity of the black hole.   Therefore, for purposes of a rough estimate, we simply consider a straight line encounter and use $b$ for \rperi.  Then we find for the energy loss in the encounter relative to the initial energy a ratio
$\Delta E/E \approx - 10^{-2.01} M_{10} \vctwo^2 b_{10}^{-1} v_3^{-4}$.
With $M_{10} = 1.7$ and $\vctwo = 1$ for NGC~1277, this gives
$\Delta E/E \approx - 10^{-1.78} b_{10}^{-1} v_3^{-4}$.
A higher velocity requires a smaller impact parameter for capture ($\Delta E/E  < -1$), because there is more  energy to be dissipated and the higher velocity inhibits dynamical friction. 
The best chance for capture involves an encounter with a relatively low velocity in comparison to the cluster velocity dispersion.   For example, capture with
$b = 10~\kpc$ requires $v < 360~\kms$ while $b = 30~\kpc$ requires $v < 270~\kms$

Let us assume that  the 9 galaxies discussed above (after excluding NGC~1275) are contained in a cubical volume of $(0.24~\Mpc)^3$,  giving a volume density of $n_\mathrm{gal} \approx 610$ galaxies per
cubic Mpc.  The mean free path is then $\lambda \approx ( f_{v} n_\mathrm{gal} \pi b^2)^{-1} \approx 5  f_{v}^{-1} \,b_{10}^{-2}~\Mpc$,
where $ f_{v}$ is the fraction of the encounters with velocity less than $v$.
The typical collision time is then $\tcoll \approx \lambda/v \approx (10^{9.7}~\yr)  f_{v}^{-1} b_{10}^{-2} v_3^{-1}$.
If  the ejected black hole spends much of its time near
the apocenter of its orbit and has a relatively low velocity, then we may estimate $f_v$ from the velocity distribution of the galaxies.
Of the nine bright galaxies in the subcluster other than NGC~1275, two have velocities within $200~\kms$ of NGC~1275
($\Delta v = +87$ and $-78~\kms$).  If this fraction applies in all three dimensions, then roughly $f_v \approx (2/9)^3 = 10^{-2.1}$ for $v = 360~\kms$, since $360/\sqrt{3} = 208$.   (A Maxwellian distribution with $\sigma = 1000~\kms$ gives $f_v = 10^{-2.4}$ for $v = 300~\kms$.)  Then with $b_{10} = 1$, we find a capture probability of about $10^{-1.6}$ in a Hubble time.  Likewise, for 
$b_{10} = 3$ and $v = 270~\kms$, we find a capture probability of about $10^{-1.1}$ in a Hubble time, where we have scaled
$f_v$ as $v^3$.  This assumes that the halo of NGC~1277 persists beyond 30~\kpc.
 
The implication of this very rough 
estimate is simply that capture of the runaway hole by a galaxy of the size of  NGC~1277 is possible. 
Several authors have discussed evidence for low velocity subclusters surrounding some BCGs  
\citep[e.g.,][and references therein]{gebhardt91}.  If such a subcluster was in existence at the time that NGC 1275 ejected its black hole, the probability of capture may be enhanced.
More generally, the ejection event may have occurred substantially earlier in the evolution of the cluster, so that estimates of the probability of capture are necessarily uncertain.
Once in orbit, dynamical friction leads to the in-spiral of the black hole to the center of the galaxy in a time
$\sim10^{8}~\yr$.  Thus, if the capture occurred more than a billion years ago,
there has been ample time for the galaxy to settle down to its current, symmetrical appearance.

\section{Discussion}
\label{sec:discuss}

The scenario outlined here is speculative, but it involves known processes.  
Any explanation of the UMBH in NGC 1277 is likely to involve exceptional events.   

Is the present day appearance of NGC 1275 consistent with the idea that it long ago formed and ejected
a UMBH?  One indicator might be the presence of an unexpectedly small black hole in the nucleus today.
Scaling linearly in $L_\mathrm{V}$ from NGC 4889 or NGC 3842, one might expect $\mbh \approx 10^{9.5}\msun$
in NGC 1275.   In contrast, \citet{wilman05} derive $\mbh \approx 10^{8.5}\msun$ from a study of the
molecular gas in the nucleus. 
 A larger value $\mbh \approx 10^{8.9}~\msun$ is derived by \citet{scharwachter13},
although these authors regard their measurement as an upper limit because of the abundance of gas in the nucleus.
These values equal or exceed the value predicted by the \mbhsigstar\ relationship \citep{wilman05}.  
However, \citet{lauer07a} argue that for the largest galaxies, \mbh\ is better predicted by stellar luminosity than by velocity dispersion,
in which case the black hole in NGC 1275 is indeed undersized.   The observed black hole may have been regenerated by means of mergers or accretion of gas following the ejection event.
Merger tree simulations by \citet{volonteri07} indicate that a dark matter halo of mass $\sim10^{12}~\msun$ will likely have undergone one or
two major mergers more recently than $z = 0.5$, likely introducing a substantial black hole that would spiral to the nucleus of the merged galaxy. Alternatively,
\citet{inoue96} find $3\times10^{10}~\msun$ of molecular gas inside a radius of 10~kpc in NGC 1275, with $6\times10^9~\msun$ contained
in a ring of radius 1.2~kpc around the nucleus, so that ample gas is available to form a massive black hole.
The physics regulating the \mbhsigstar\ relationship is not well established, and some feedback process might lead to
regrowth of a black hole to a limiting value similar to the one given by the normal \mbhsigstar\ relationship \citep[e.g.,][]{dimatteo05}.  

Many large elliptical galaxies have 
cores in which the density profile increases toward the center more gently than in galaxies with a central cusp.    
One explanation of these cores is scouring by
a binary SMBH during its in-spiral to the nucleus \citep{milos01}.   \citet{lauer07a,lauer07b} find that cores are
prevalent in brighter ellipticals, with $M_\mathrm{V} <  -21$.  The mass deficit involved in these cores
 is of order the central black hole mass.
If NGC 1275 formed and ejected a UMBH, then a substantial core might be expected,
with a radius $\sim 1~\kpc$ and a mass deficit out of proportion to the
mass of the current black hole.  
[\citet{postman12} find a large core in A2261-BCG and suggest that it may have ejected a UMBH.]
Optical and ultraviolet surface brightness profiles for NGC~1275 by \citet{marcum01}  show no break down to a radius $\sim 0.3~\kpc$.  However, 
\citet{lauer07a}  find considerable scatter in core mass and radius relative to \mbh.  
Furthermore, \citet{postman12} note that
cores that have lost their central black hole may quickly be filled in by inspiralling nuclei from captured galaxies.

Is the present day appearance of NGC 1277 consistent with this scenario?    One possible difficulty is the lack of a classical bulge.
VB12 report only a psuedo-bulge in NGC 1277 having 24\%\ of the light, the rest being in a flattened disk.
Simulations of the capture are needed to determine whether a merger with a black hole having 14\%\ of the stellar mass of the
galaxy can avoid forming a bulge and disrupting the disk. 
The merger could add a substantial amount of angular momentum to the galaxy.  
VB12 find a rotational velocity 
$\vrot \approx 250~\kms$ for NGC 1277, and an effective radius for the starlight of $R_\mathrm{e} = 1.6~\kpc$.   If the black hole approaches
with an impact parameter $b$ and a relative velocity \vrel, then its angular momentum as a fraction of the current rotational angular
momentum of the galaxy is 
\begin{equation}
J_\mathrm{BH}/J_\mathrm{gal} \approx 10^{0.2} b_{10} M_{10} v_{\mathrm rel, 3}, 
\label{eq:angmom}
\end{equation}
where $b_{10} = b/10~\kpc$, $M_{10} = \mbh/10^{10}~\msun$, and $v_{\mathrm rel, 3} = \vrel/10^3~\kms$.
The merger could have contributed substantially to the current rotation of the galaxy.

Black hole migration in galaxy groups and clusters may occur with some frequency.  \citet{volonteri07} discusses the ejection of black holes following galaxy mergers in cosmological simulations and illustrates how this can contribute downward scatter in the \mbh\ - bulge relationships.   As noted by \citet{blecha11}, runaway black holes may stand a good chance to be reincorporated into adoptive galaxies.  This can contribute upward scatter to the black hole - bulge relationship when large black holes fall into relatively small galaxies.  NGC 1277 may be an extreme example.

If the UMBH in NGC 1277 grew by accretion, then the average growth rate over the Hubble time is $\mdot \approx 1~\msunyr$.  Luminous accretion at this rate produces a luminosity $L \approx 10^{46}~\ergps$.  The relative proximity of NGC 1277 to the Milky Way together with the observed space density of luminous quasars places constraints on the growth of the UMBH by accretion of gas  \citep{fabian13}.  These requirements are eased if the UMBH reached its final size by means of mergers.

Our scenario requires a combination of seemingly unlikely events.  NGC 1275 must have produced an exceptionally massive black hole though a merger of two fairly equal mass black holes, themselves already comparable to the largest black holes in nearby BCGs.  The merger must have had a favorable spin-orbit configuration leading to a large recoil velocity.  And the runaway black hole must have been captured by NGC 1277, an unusual galaxy in terms of its velocity dispersion and compactness.   However,  rare events do occur,  and it is important to consider all possibilities.  Confirmation of the black hole migration scenario would have significant implications for the evolution of galaxies and for our understanding of general relativity in the strong field limit.

\acknowledgments
We thank the referee for valuable suggestions that improved the manuscript and Carlos Lousto for helpful discussions.  We acknowledge the usage of the HyperLeda database (http://leda.univ-lyon1.fr).  This research has made use of the NASA/IPAC Extragalactic Database (NED) which is operated by the Jet Propulsion Laboratory, California Institute of Technology, under contract with the National Aeronautics and Space Administration (http://ned.ipac.caltech.edu).

\end{document}